\pdfoutput=1

\documentclass[10pt,journal,compsoc]{IEEEtran}

\usepackage{biblatex}  
\ifCLASSINFOpdf
\else
\fi
\usepackage{biblatex} 
\usepackage{multirow}
\usepackage{amsmath}
\usepackage{booktabs} % 用于漂亮的表格线条
\usepackage{graphicx}

\usepackage{amsthm}
\usepackage{graphicx} % 自适应调整表格
\usepackage{algorithm}

\usepackage{algpseudocode}

\usepackage{subfigure}

\usepackage{amssymb}

\usepackage{color}
\usepackage{bm}

\newcommand{\degree}{^\circ}

\usepackage{url}

\begin{document}
\title{The establishment of static digital humans and the integration with spinal models}

\author{Fujiao~Ju$^{1}$, Yuxuan~Wang$^{1}$, Shuo~Wang$^{2}$, Chengyin~Wang$^{3}$, Yinbo~Chen$^{3}$, Jianfeng~Li$^{3}$, Mingjie~Dong$^{3*}$, Bin~Fang$^{4*}$, Qianyu~Zhuang$^{5*}$
\thanks{$^{1}$Fujiao~Ju, Yuxuan~Wang are with the College of Computer Science, Beijing University of Technology, Beijing 100124, P.R. China.}
\thanks{$^{2}$Shuo~Wang is with the Department of Engineering Physics, Tsinghua University, Beijing 100084, P.R. China.}
\thanks{$^{3}$Chengyin~Wang, Yinbo~Chen, Jianfeng~Li, and Mingjie~Dong (dongmj@bjut.edu.cn) are with the Beijing Key Laboratory of Advanced Manufacturing Technology, College of Mechanical \& Energy Engineering, Beijing University of Technology, Beijing, 100124, P.R. China.}
\thanks{$^{4}$Bin~Fang (fangbin1120@bupt.edu.cn) is with the Department of Artificial Intelligence, Beijing University of Posts and Telecommunications, Beijing, 100876, P.R. China.}
\thanks{$^{5}$Qianyu~Zhuang (zhuangqianyu@hotmail.com) is with the Department of Orthopedics, Peking Union Medical College Hospital, Beijing, 100730, P.R. China.}
\thanks{This work was supported in part by the  Beijing Natural Science Foundation under Grant No. L222012, in part by the open project of the key laboratories and engineering technology research centers in the field of rehabilitation under the Ministry of Civil Affairs under Grant No. 102118170090010009004, in part by the Beijing Nova Program under Grant No. 20230484488, and in part by the National High Level Hospital Clinical Research Funding under Grant No. 2022-PUMCH-B-002.}
\thanks{*Corresponding authors (Mingjie~Dong; Bin~Fang; Qianyu~Zhuang).}
}

\markboth{Journal of \LaTeX\ Class Files,~Vol.~XX, No.~XX, December~2024}%
{Shell \MakeLowercase{\textit{et al.}}: Bare Demo of IEEEtran.cls for IEEE Journals}



\IEEEtitleabstractindextext{%
\begin{abstract}

Adolescent idiopathic scoliosis (AIS) is a common spinal deformity that has a profound impact on health and quality of life. Although conventional imaging techniques, e.g., X-rays, computed tomography (CT), and magnetic resonance imaging (MRIs), provide static views of the spine, they are limited in capturing the dynamic changes in the spine and its interactions with overall body motion. Therefore, it has become particularly important to develop new techniques to fill these gaps. Dynamic digital human modeling is a major advancement in digital medicine that allows the spine to be viewed in three dimensions (3D) as it changes during daily activities, helping clinicians identify deformities that may be overlooked in static imaging. Although dynamic modeling has great potential, constructing an accurate static digital human model is an essential first step to achieve high-precision simulations. In this study, we focus on constructing an accurate static digital human model integrating the spine, which is crucial for subsequent dynamic digital human research in AIS. Initially, human point cloud data is generated by combining the 3D Gaussian method with the Skinned Multi-Person Linear (SMPL) from the patient's multi-view images. Subsequently, a standard skeletal model is fitted to the generated human model. Next, the real spine model reconstructed from CT images is aligned with the standard skeletal model. The resulting personalized spine model has been validated using X-ray data from six patients with AIS, with Cobb angles (measuring the severity of the scoliosis) as evaluation metrics. The results show that the error of the model was within $1\degree$ of the actual measurements. This study provides an important method for constructing digital humans.
\end{abstract}

}

\maketitle

\IEEEdisplaynontitleabstractindextext

% \fi
\IEEEpeerreviewmaketitle

%%%%%%%%%%%%%%%%%%%%%%%%%%%%%%%%%%%%%%%%%%%%%%%%%%%%%%%%%%%%%%%%%%%%%%%%%%%%%%%%%%%%%


%%%%%%%%%%%%%%%%%%%%%%%%%%%%%%%%%%%%%%%%%%%%%%%%%%%%%%%
%%%%%%%%%%%%%%%%%%%%%%%%%%%%%%%%%%%%%%%%%%%%%%%%%%%%%%%
\section{Introduction}
\par Adolescent idiopathic scoliosis (AIS), a spinal deformity that commonly manifests in adolescents, is distinguished by an abnormal side-to-side curvature. This condition not only modifies physical appearance but also exerts a significant influence on health and overall quality of life. According to Weinstein et al. \cite{weinstein2008adolescent}, severe scoliosis can compress the lungs, leading to diminished lung capacity and shortness of breath during physical exertion. Additionally, thoracic deformities may impair heart function, resulting in symptoms such as fatigue and dizziness. The asymmetric curvature disrupts muscle balance, which, as noted by Chin and Kingsley \cite{CChin-Kingsley-R2001}, restricts mobility and the performance of daily activities. Furthermore, AIS can adversely impact mental well-being by inducing feelings of anxiety, low self-esteem, and social challenges. These physical and psychological impacts underscore the necessity of accurate research and effective treatment for AIS.

\par Traditional medical imaging techniques, such as X-rays, computed tomography (CT), and magnetic resonance imaging (MRI), have been widely used for diagnosing and evaluating spinal conditions like AIS. However, these methods have notable limitations, particularly in capturing the dynamic or weight-bearing state of the spine. For instance, X-rays have long been considered the gold standard for visualizing spinal deformities, including scoliosis. However, conventional X-ray imaging provides limited insights into the full range of spinal motion or the effects of weight-bearing on the spine. Moreover, X-ray imaging exposes patients to ionizing radiation, which poses health risks. Studies, such as the systematic review by Luan et al. \cite{luan2020cancer}, have shown that repeated X-ray exposure in scoliosis patients can result in significant cumulative radiation doses, increasing risks of cancer and other adverse outcomes. Larson et al.~\cite{larson2019radiation} also highlighted concerns about the risks of ionizing radiation, particularly for young individuals requiring frequent imaging during treatment.
Additionally, traditional imaging methods struggle to fully capture the 3D complexity of spinal deformities. Scoliosis, for example, often involves a combination of lateral curvature, rotation, and sagittal misalignment, which cannot be adequately represented by two-dimensional imaging techniques. This limitation hinders accurate assessment of deformity severity and complicates the development of effective treatment strategies. As emphasized by Illés et al. \cite{illes2019third}, scoliosis is inherently a 3D deformity, and traditional methods often neglect the axial plane, which is crucial for understanding the full extent of the condition. CT scans are typically conducted in a fixed supine position, producing static images of the spine in a neutral, non-functional posture. This limitation is significant, as static images fail to represent the spine's actual condition during daily activities when subjected to weight-bearing and dynamic forces. In addition, CT imaging primarily focuses on the trunk, often neglecting to capture full-body posture, including limb positions. This omission can be critical, as understanding overall musculoskeletal alignment and posture is essential for comprehensive patient assessment. 

\par In order to gain a more comprehensive understanding of the impacts that AIS exerts on the spine, it is crucial to capture both the static morphological features and explore the dynamic performance of the spine during motion. Traditional gait analysis methods, which are commonly used to study spinal dynamics, have notable limitations. For instance, Shull et al. \cite{shull2016magneto} highlighted that these methods often required patients to wear hardware devices, making them inconvenient for home use. Frigo et al. \cite{frigo2003upper} further noted that such devices might interfere with the naturalness of patient movements, affecting the accuracy of the analysis. Similarly, Mahaudens and Philippe \cite{Mahaudens-Philippe2009} discussed how these methods might introduce variability in motion studies, while Nishida and Mitsuhiro \cite{Nishida-Mitsuhiro2017} emphasized the challenge of integrating these systems into everyday clinical practice.
In contrast, digital human technology based on visual composition offers significant advantages. It non-invasively captures dynamic spinal information without requiring patients to wear additional equipment. Using advanced visual recognition and analysis algorithms, this technology accurately records and analyzes spinal movements and postures during various activities, providing a more detailed and precise understanding of the spine's dynamic characteristics. Additionally, it can simulate surgical outcomes and predict disease progression, establishing a more reliable foundation for personalized treatment plans. This technology holds great promise for advancing the research and treatment of AIS.

\par Before delving into the dynamic digital human research, constructing an accurate static digital human model is essential. A reliable static model provides a foundational understanding of the spine's condition. Without it, subsequent dynamic analyses may lack precision and a solid basis. Therefore, this study aims to create an accurate digital human model that incorporates the unique spinal characteristics of patients. 
\par Specifically, we generate human point cloud data by combining the 3D Gaussian method with the Skinned Multi-Person Linear (SMPL) model from the patient’s multi-view images and fit a standard skeletal model to the generated human model.  The real spine model, reconstructed from CT images, is aligned with the standard skeletal model through a feature point registration method. To ensure the accuracy of the model, an As-Rigid-As-Possible (ARAP) algorithm \cite{levi2014smooth} is applied for final optimization and fusion. This digital human model incorporates the unique spinal features of patients and lays a solid foundation for dynamic digital human research. By accurately capturing the spinal structure of individuals, this comprehensive method constructs a high-quality static digital human model integrating the spine, which serves as an important prerequisite for simulating and exploring the spinal behavior during dynamic weight-bearing activities.

In summary, our contributions include the following.
\begin{itemize}
\item \textbf{A multimodal dataset}: We synchronously collect multi-view images and CT data from the patients, providing a foundation for using the multi-view images for human body reconstruction and the CT data for constructing the real spine model.
\item \textbf{A novel static digital human model for AIS}: We develop an innovative static digital human model that incorporates the unique spinal characteristics of AIS patients.
\item \textbf{Precise registration between the real and standardized models}: We employ a feature-point coarse registration combined with the ARAP fine registration method to align the real spine model, reconstructed from CT images, with the standardized skeletal model.
\item \textbf{An effective tool for clinical applications}: We develop and validate a precisely aligned 3D spine model in clinical diagnosis and surgical planning for AIS patients. This model not only lays a solid foundation for dynamic digital human research but also enables clinicians to visually observe the relationship between the 3D spine model and human posture. This helps clinicians in formulating personalized treatment plans, improving the precision and success rate of surgeries.
\end{itemize}

\begin{figure*}
	\centering
     \includegraphics[width=\textwidth]{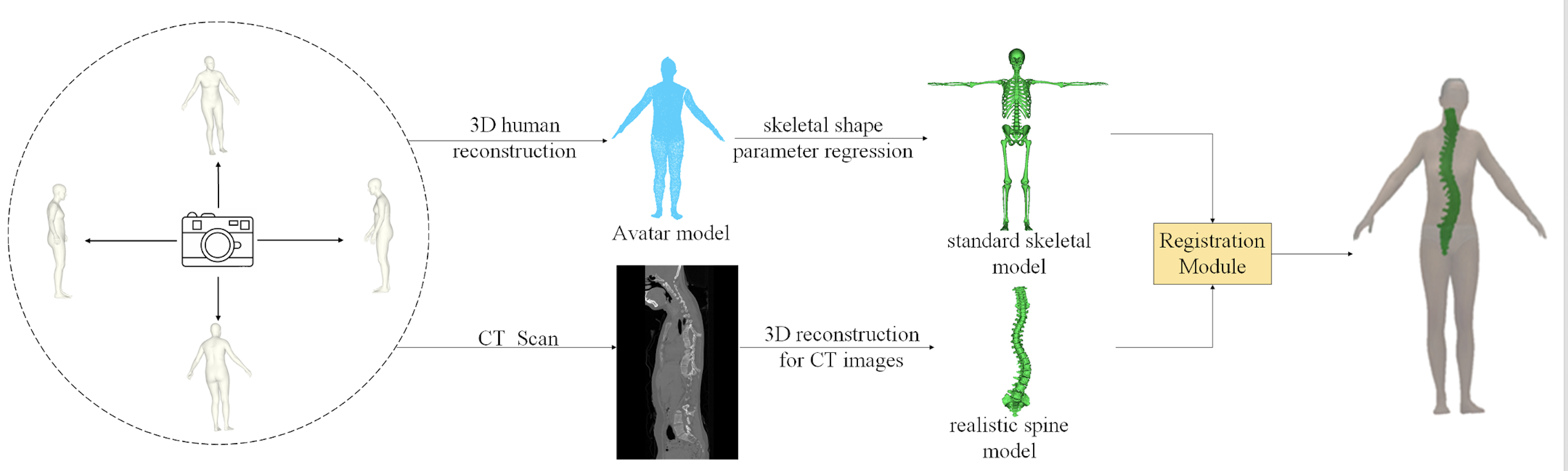}
	\caption{The framework of the proposed method.}
    \label{framework}
\end{figure*}

%%%%%%%%%%%%%%%%%%%%%%%%%%%%%%%%%%%%%%%
\section{Related Works} \label{Sec:related}
\subsection{Human Model}
In current researches on dynamic spine modeling, the parametric method of human body models plays an important role. The SMPL model is a widely used parametric human body model that effectively describes and generates 3D human shapes through a small number of parameters. This model uses the Linear Blended Skin (LBS) algorithm to combine pose parameters with shape parameters, thereby achieving a flexible representation of human morphology. This makes SMPL widely used in various computer vision and graphics applications, especially in tasks such as human pose estimation and motion capture \cite{loper2023smpl}.
In addition to this classic model of SMPL, in recent years, the field of parametric human body modeling has ushered in a new wave. In particular, Neural Radiance Fields (NeRF) \cite{mildenhall2021nerf} and its innovative applications on human subjects have pushed this field to a new height. Human NeRF methods proposed by Chung-Yi Weng et al. \cite{weng2022humannerf}, for example, are remarkable for their extraordinary ability to directly encode human geometry and appearance into neural networks and achieve realistic rendering. However, while these methods pursue extreme realism, they also face the dual challenges of computational efficiency and training complexity.
\par It is worth mentioning that in the vast field of parametric human body modeling, there is another method that is also worthy of attention, that is, 3D Gaussian human body modeling, which uses the characteristics of Gaussian distribution to parametrically represent human shape. By dividing the human body into multiple parts and describing each part with a 3D Gaussian distribution, a fine depiction of human shape can be achieved. For example, GauHuman \cite{hu2024gauhuman}, as an efficient 3D human body modeling algorithm, takes its articulated Gaussian splatting representation as the core and is specially designed for fast training and rendering. This method encodes Gaussian splatting in the canonical space and maps it to the pose space. Using the LBS weights of the SMPL model as a starting point, it predicts the corresponding LBS weight offsets through a multilayer perceptron (MLP), thereby achieving fast and accurate modeling of human shape. Similar to Human NeRF, while pursuing efficient modeling, GauHuman also fully utilizes the characteristics of 3D Gaussian distribution, enabling it to quickly build a 3D human body model in just one to two minutes.

\subsection{Spine Model}
Currently, approaches to spine modeling have focused on multibody dynamics and finite element method (FEM). For example, Lu et al. used Mimics and Ansys software to establish a 3D finite element model of the thoracolumbar segment T12 - L2 based on the CT scan data of the T12 - L2 vertebral body of a 40-year-old healthy female volunteer~\cite{lu2022establishment}. They verified the effectiveness of the model by simulating multiple working conditions and analyzing the displacement and stress distribution of the vertebral body and intervertebral disc. Similarly, Ibrahim El Bojairami et al. developed and validated a 3D comprehensive finite element spine model that includes multiple tissues and physiological effects \cite{el2020development}. They used a new meshing technique to improve computational efficiency and proved its reliability through various verifications, providing an effective tool for spine-related research. However, although 3D FEM has the advantages of simplicity, speed and economy, and can simulate and calculate the mechanical properties of various materials under complex conditions, and the experiment has the advantages of being easy to adjust and repeatable, FEM still has limitations. For example, it is difficult to accurately simulate certain nonlinear properties, and it is susceptible to human influence when dealing with the parameters of the model. In addition, it is often difficult to combine the FEM with the human body's action postures in biomechanical analysis. In addition, related research also has limitations such as limited simulation range, only using finite element analysis software, and failing to effectively simulate some human anatomical structure \cite{naoum2021finite}.
\par We can attempt to address these issues by applying computer vision methods to develop a technique for constructing a human spine model. In \cite{aubert2019toward}, the authors constructed 3D spine by using CNN to fit a statistical spine model to images, which demonstrated good performance in terms of landmark location accuracy and clinical parameter extraction, with a short reconstruction time, thereby providing valuable support for 3D measurements in clinical practice. Chen et al. proposed a generative adversarial network (GAN) framework called ReVerteR \cite{chen2024automatic}, which aimed to automatically reconstruct a 3D model of the spine from orthogonal biplane X-ray maps. The framework utilized digitally reconstructed X-ray (DRR) images from coronal and sagittal planes for 3D reconstruction and developed a deep learning-driven transformation module to improve the accuracy of the reconstruction. In addition, they introduced an automatic center-of-mass annotation module, which enhanced the overall performance of the 3D reconstruction process.
In~\cite{chen2023bx2s}, the BX2S-Net deep learning framework was developed for reconstructing 3D spine structures from biplane X-ray maps. The network takes individual vertebrae in anterior-posterior (AP) and lateral (LAT) views as input and output 3D models in voxel grid format. BX2S-Net employs an incremental decoding process combined with feature fusion and attention mechanisms to optimize the 3D reconstruction results.
\par The above studies demonstrate recent advances in the field of reconstructing spine models from X-rays, employing different deep learning approaches and providing high-quality 3D spine models, which provide important references in addressing the limitations of traditional methods. 

\begin{figure*}
	\centering
     \includegraphics[width=\textwidth]{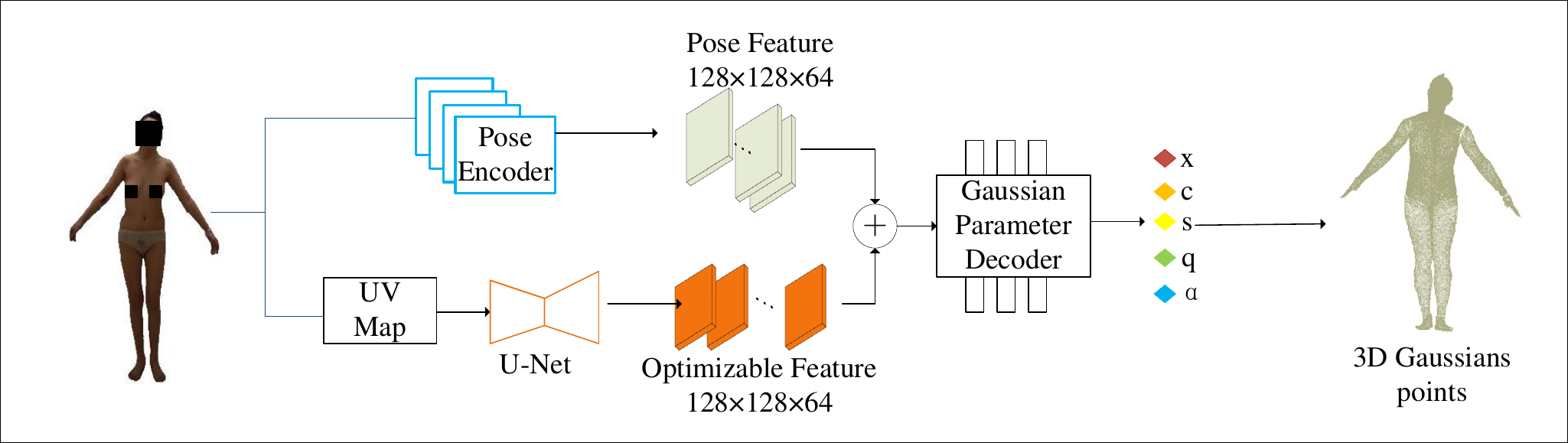}
	\caption{In the given frame, sampling operations are carried out on the human body surface to form a point cloud, and at the same time, the corresponding positions of these points are marked in the UV position map. After that, the UV position map is fed into the pose encoder to generate the corresponding pose features. During this period, the optimizable feature tensor is precisely aligned with the pose features in a specific way, with the aim of more effectively capturing the overall appearance of the human body. These aligned features are input into the Gaussian parameter decoder, which can predict the offset $\delta x$, color c and scale s of each point. And these predicted results, together with the fixed rotation q and opacity $\alpha$, jointly form an animatable 3D Gaussian distribution in the canonical space.}
    \label{network}
\end{figure*}

%%%%%%%%%%%%%%%%%%%%%%%%%%%%%%%%%%%%%%%%%%%%%%%%%%%%%
%%%%%%%%%%%%%%%%%%%%%%%%%%%%%%%%
\section{Methodogy} \label{Sec:method}
\par In this study, we constructed a systematic method around multimodal data to achieve the establishment of a static digital human body model and the integration of the spine model. The overall process framework is as in Fig.~\ref{framework}. Firstly, we synchronously collect the multi-view images and CT data of the patient, which lays a foundation for subsequent utilization of these data to conduct human body reconstruction and construction of the real spine model respectively. Then, the GS method combined with the SMPL model is used to convert the multi-view image data into human point cloud data, and the bone landmark point parameters are further calculated through constraints to obtain the standard skeletal model. Finally, the feature point coarse registration and ARAP registration methods are used to register and optimize the real spine model constructed from the CT image and the standard skeletal model, and a static digital human model with precise registration is obtained.
\subsection{GS method to generate human body point clouds}
\par 3D Gaussian expansion is a point-based scene representation technique that enables high-quality real-time rendering \cite{kerbl20233d}. This scene representation is parameterized by a set of static 3D Gaussians. Each static 3D Gaussian has the following parameters: 3D center position $x\in R^3$, opacity $\alpha \in R$ , 3D rotation in quaternion form $q\in R^4$, and 3D scale factor $s\in R^3$. Through these attributes, we can generate rendered images from any viewpoint. 

\par To extend the above representation to human head modeling, we integrate it with the SMPL model as in~(\ref{Splatting}): 
\begin{equation}
    G(\beta,\theta,D,P)=\text{Splatting}(W(D,J(\beta),\theta,\omega,P),
    \label{Splatting}
\end{equation}
where, $G(\cdot)$  represents the rendered image, $\text{Splatting}(\cdot)$ represents the process of rendering 3D Gaussians from any angle, $W(\cdot)$ is the standard linear blend skinning function used to place 3D Gaussians, $D=T(\beta)+dT$ represents the position of the 3D Gaussian in canonical space, formed by adding the corrected point displacement $dT$ to the surface of the template mesh $T(\beta)$, $P$ represents other attributes of the 3D Gaussian excluding the position, $\beta$ and $\theta$ are shape and pose parameters, and $J(\beta)$ outputs the 3D joint positions. It should be noted that we propagate the skinning weights  from the vertices of the SMPL model to the nearest 3D Gaussian. Through the proposed representation method, we can transfer these canonical 3D Gaussians to the motion space for free-viewpoint rendering.
Based on the proposed Gaussian model, the human appearance is determined by the point displacement $dT$ and the attribute $P$. Modeling the dynamic human appearance can be regarded as estimating these dynamic attributes. 
\par In the process of obtaining human body point cloud data based on the 3D Gaussian model, a series of key technical components work in synergy. We generate the UV map (UV herein refers to the abbreviation of the u and v texture mapping coordinates, which is similar to the X, Y, Z axes of the spatial model and is designed to define the position information of each point on the image; These points are interrelated with the 3D model to determine the specific position of the surface texture mapping )  using the SMPL model and input it into the Pose Encoder with a specific architecture for feature extraction. Subsequently, combined with the optimized feature tensor, the final 3D Gaussian point cloud is generated through the Gaussian Parameter Decoder. Among them, the relationship between each component and the data processing flow is shown in Fig.~\ref{network}, which clearly illustrates how input data is gradually transformed into human body point cloud data, providing an intuitive architecture diagram for understanding this complex process.

To model the dynamic human appearance under various postures, a dynamic appearance network and an optimizable feature tensor are used to predict these pose-dependent attributes of the 3D Gaussian human body. The dynamic appearance network aims to learn the mapping from the two-dimensional manifold representing the underlying human shape to the dynamic attributes of the 3D Gaussian human body, as follows: $f_\phi :S^2 \in R^3 \longrightarrow R^7$.
Luiten et al. \cite{luiten2023dynamic} introduced a framework that effectively maped surface manifolds to dynamic features, enabling the detailed reconstruction of human body appearance under varying poses. Building on this foundation, Hu et al. \cite{hu2024gaussianavatar} proposed GaussianAvatar, a novel approach that integrated dynamic networks with Gaussian point cloud representations, further improving the accuracy of obtaining human body Gaussian point cloud data from multi-view images of the patient.
These methods realize the process of obtaining human body Gaussian point cloud data from the multi-view images, combining the advancements proposed in these works to address the challenges of dynamic human modeling.

%%%%%%%%%%%%%%%%%%%%%%%%%%%%%%%%%%%%%%%%%%%%%%%%%%%%%%%%%%%%%%%%%%%%%

\subsection{Obtain standard skeletal model}
\par
\begin{figure}
  \centering
  \includegraphics[width=7cm,page={1}]{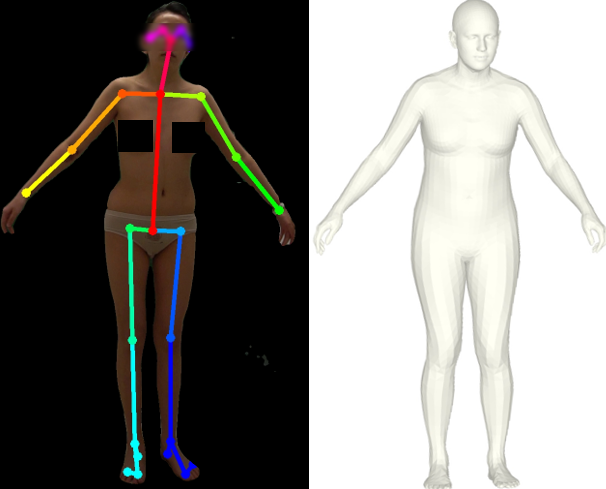}
  \caption{SMPL 3D model and skeletal landmarks}
  \label{SMPL}
\end{figure}

 Current 3D human pose estimation methods approximate the body's motion structure using a simplified skeleton composed of a few segments connected by spherical joints. However, these simplified skeletons are not suitable for modeling digital humans with scoliosis. To simulate scoliosis more accurately, we project the 3D pose, generated using the SMPL model and camera parameters from Section 3.1, onto the 2D image plane. In the projected 2D image, specific joints or skeletal landmarks such as the head, shoulders, elbows, and wrists are extracted as 2D skeletal landmarks, as shown in Fig.~\ref{SMPL}. These skeletal landmarks form the initial spatial configuration of the standard skeletal model. Keller et al. \cite{keller2023skin} described a process for utilizing the SMPL model to refine and reconstruct 3D human skeletal structures, demonstrating its flexibility in various applications. Similarly, Keller and Osborne \cite{keller2022osso} proposed the Osso framework, which further enhanced the fitting of standard skeletal models to SMPL-generated 3D poses by incorporating joint-specific optimizations. Building on these methods, the standard skeletal model is fitted to the SMPL 3D model to better represent the spatial configuration and alignment of individuals with scoliosis.

\par In the adjustment process of the skeletal model, although the SMPL model provides detailed 3D geometric information for the estimation of human pose, it is not fully closed in its topological structure, and certain local areas may have holes or discontinuous boundaries. As a result, directly using the SMPL model for skeletal fitting and simulation may not fully ensure the integrity and accuracy of the surface, especially when high-precision interaction between bones and skin is required, potentially affecting the precise positioning of bones.

\par To address these issues and generate a more accurate, topologically simple surface, we chose to convert the SMPL model into a genus-0 surface. By generating a closed and hole-free genus-0 surface, we can ensure that the surface has good manifold characteristics. This surface serves as a more stable basis for skeletal fitting. It is a smooth, continuous 2D manifold in a 3D space (since a 2D manifold is usually considered within a higher-dimensional embedding space). This surface excludes regions such as the interior cavities of the ribs or small openings between the pelvis or between the ulna and radius. This construction helps avoid potential geometric discontinuities present in models like SMPL and creates more favorable topological circumstances for subsequent skeletal fitting and physical simulation processes.
\begin{figure}
  \centering
  \subfigure[The front of the skin surface]{  % 第一个子图，xx 为第一个字体的名称
     \includegraphics[width=0.22\textwidth]{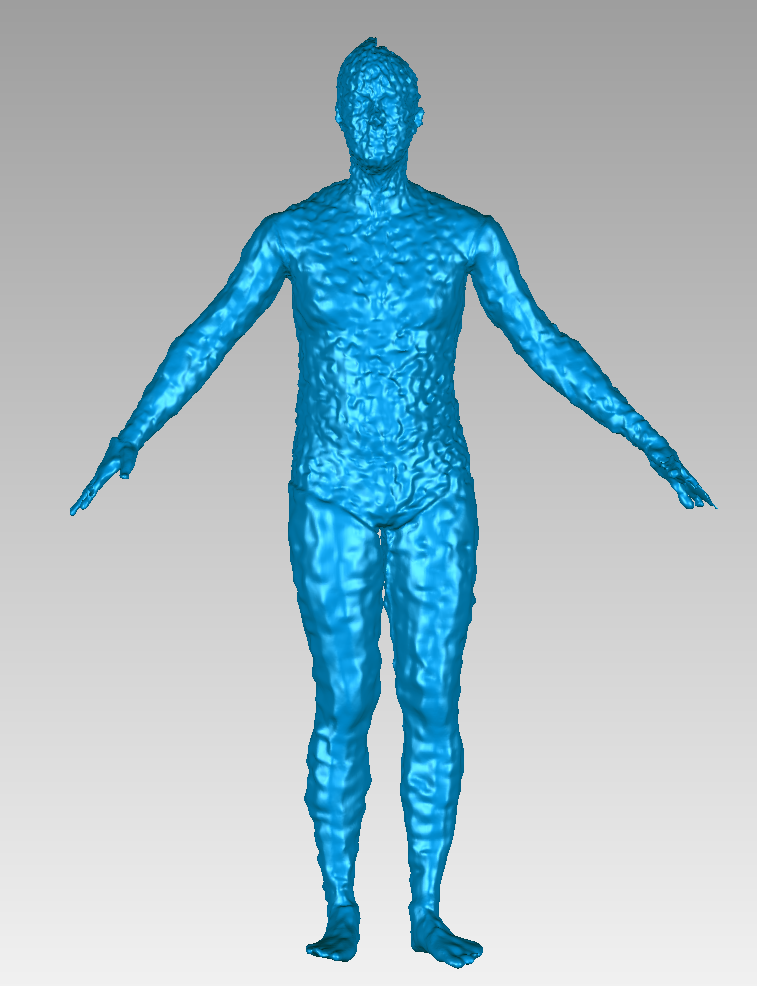}  % 图片路径和大小
   }
  \subfigure[The back of the skin surface]{  % 第二个子图
    \includegraphics[width=0.212\textwidth]{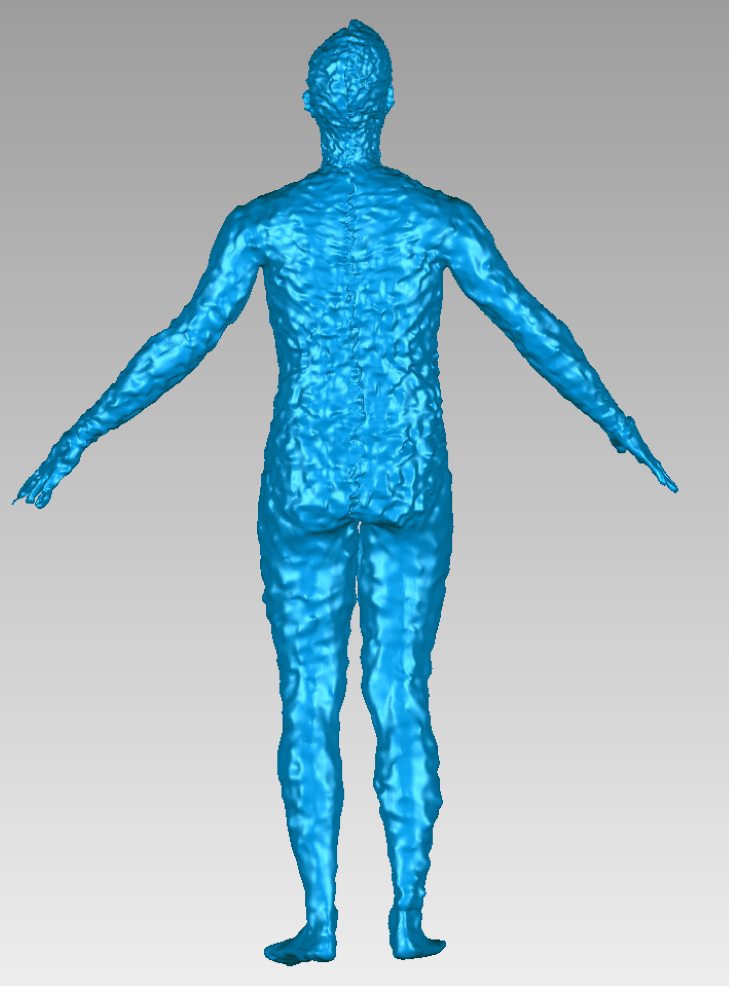}
  }
  \caption{Skin surface $ M $ }    % 整个大图的标题
  \label{Surface}
\end{figure}
\par To generate a more accurate watertight surface of genus-0, we firstly calculate the normals of the vertex of the SMPL model, obtaining the geometric directional information for each vertex. Using these normals, combined with the vertex positions, we apply the Poisson Reconstruction algorithm to successfully generate a smooth, watertight surface. This process ensures that the generated surface has good manifold properties and does not contain irregular boundaries or holes. Finally, the generated watertight genus-0 surface wraps around the skeleton, referred to as the skin surface $M$, as shown in Fig.~\ref{Surface}. This enclosing surface is a smooth, impermeable 2D manifold surface, excluding small holes within the ribcage, pelvis, or between the ulna and radius.
\par After that, we generate the skeletal 3D model B based on the skin surface $M$ as in~(\ref{energy}). At first, we set $X=M$, and then use non-linear least squares to minimize the energy, which consists of a fitting term and a regularization term. The fitting term attracts the surface $X$ to the skeleton wrapping surface $W$, while the regularization term prevents $X$ from deforming physically unreasonable from its initial state $\hat{X}=M$.
\begin{equation}
    B=\arg \min_X {w_{f_{it}}\cdot E_{f_{it}}(X,W)+w_{reg}\cdot E_{reg}(X,\hat{X})} 
    \label{energy}
\end{equation}
\par The regularization is expressed as a discrete bending energy, which penalizes changes in the mean curvature,as in~(\ref{regularization}):
\begin{equation}
    E_{reg}(X,\hat{X})=\sum_{x_i\in X}A_i||\Delta x_i-R_i\Delta \hat{x}_i||^2,
    \label{regularization}
\end{equation}
where $x_i$ and $\hat{x}_i$  represent the vertex positions of the deformed surface X and the initial surface $\hat{X}$, respectively. $R_i \in SO(3)$ is the optimal rotation matrix that aligns the Laplacians $\Delta x_i$ and $\Delta \hat{x}_i$, and $A_i$ represents the Voronoi area.

The fitting term penalizes the squared distance between the vertices $x_i \in X$ and the target positions $t_i\in W$, as in~(\ref{eq4}).
\begin{equation}
    E_{f_{it}}(X,W)=\sum_{x_i\in X}w_iA_i||x_i-t_i||^2
    \label{eq4}
\end{equation}

\begin{figure}
   \centering
   \subfigure[SMPL model with skeletal landmarks]{  % 第一个子图，xx 为第一个字体的名称
      \includegraphics[width=3.8cm]{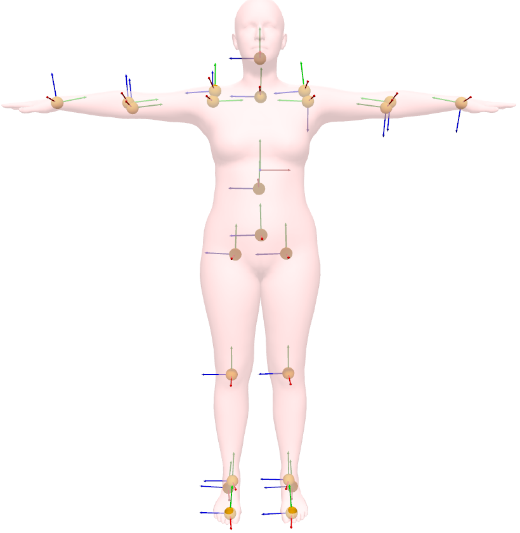}  % 图片路径和大小
    }
   \quad
   \subfigure[A fitted and generated standard skeleton model ]{  % 第二个子图
      \includegraphics[width=3.8cm]{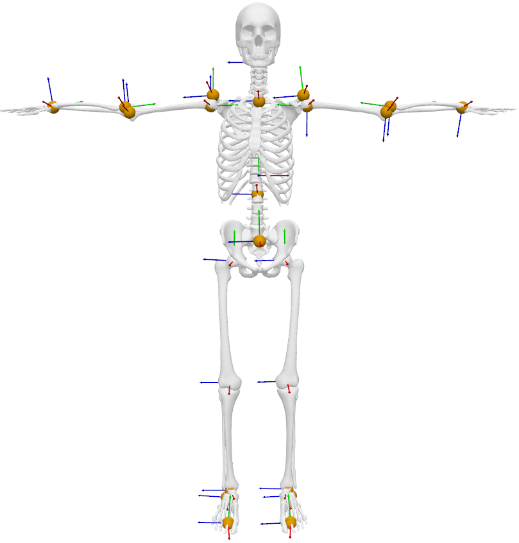}
    }
    \caption{A fitted and standard skeleton model based on the SMPL 3D model }  
    \label{fitting}
\end{figure}

The target positions $t_i$ are points on the skeleton wrapping surface $W$ and come in three types: closest point correspondence, fixed correspondence or collision target. The weights $w_i$ are determined by the type of target position $t_i$ (0.1 for the closest point correspondence, 1 for fixed correspondence, and 100 for collision target). By minimizing the energy in~(\ref{energy}), the final skeletal 3D model B is obtained, as shown in Fig.~\ref{fitting}.
 
%%%%%%%%%%%%%%%%%%%%%%%%%%%%%%%%%%%%%%%%%%%%%%%%%%%%%%%%%%%%%%%%%%%%%
\subsection{Spine registration of CT images}
To illustrate how to achieve alignment between intraoperative inspection images and preoperative medical imaging using bony landmarks, this implementation example focuses on the spine. We employ a feature point coarse alignment and ARAP registration method to align and optimize the overall spinal model.

\par
\begin{figure}
    \centering
    \includegraphics[width=0.5\textwidth]{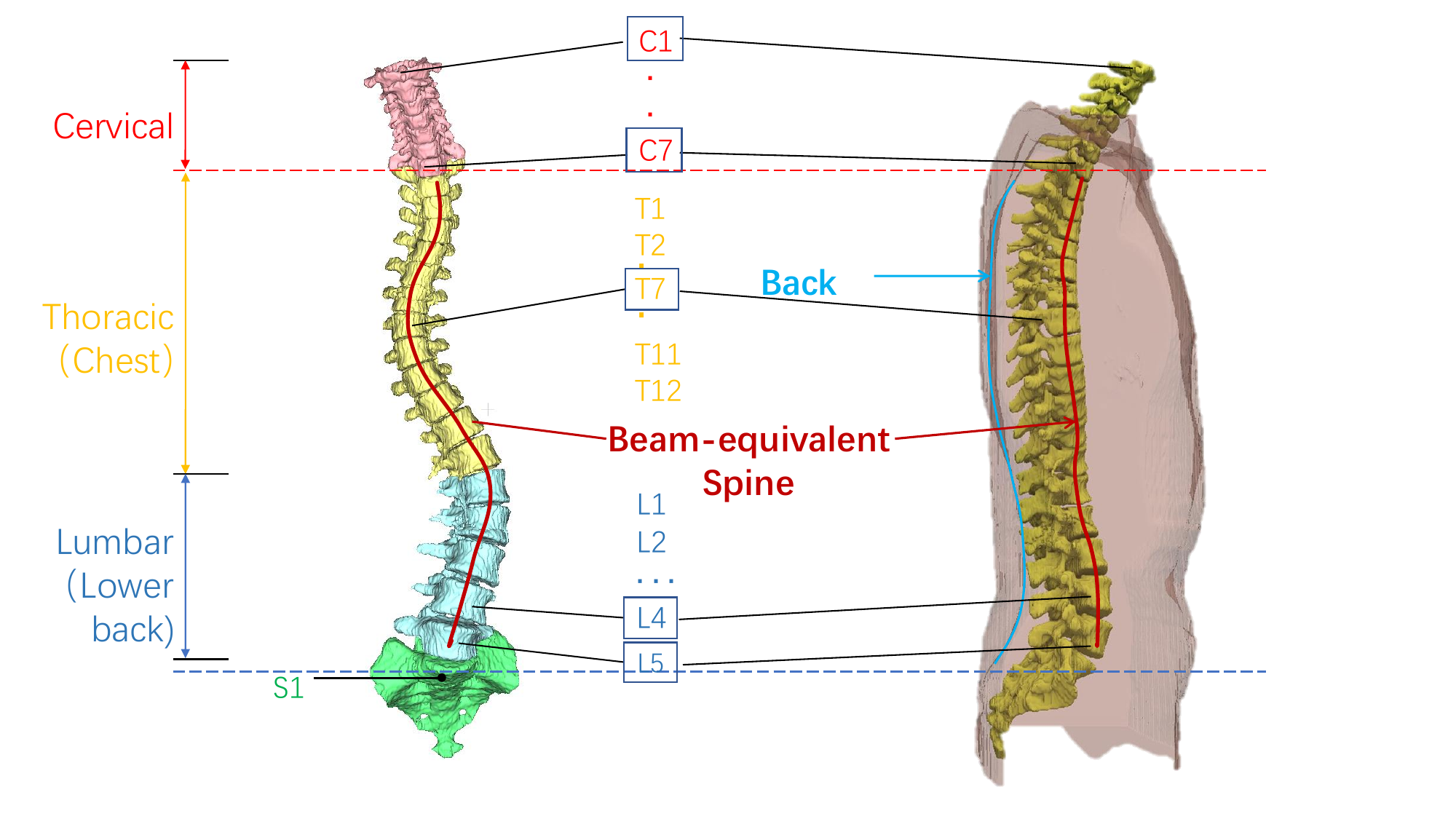}
    \caption{Selection of bone landmark points}
    \label{marks}
\end{figure}

 Firstly, we select the centroid of the following bone landmarks from the reconstructed spinal model of preoperative medical imaging (as shown in Fig.~\ref{marks}): the first cervical vertebra (C1), the seventh cervical vertebra (C7) \cite{wang2017cervical}, the seventh thoracic vertebra (T7), the fourth lumbar vertebra (L4) and the fifth lumbar vertebra (L5), denoted as $S=\{C_1,C_7,T_7,L_4,L_5\}$. These bony landmarks are chosen because they are frequently selected in clinical surgeries, easily identifiable, and play a critical role in spinal location. For example, C7 is the lowest cervical vertebra, characterized by a prominent spinous process known as the vertebra prominens.
\par In section 3.2, we similarly select the aforementioned bony landmarks on the standard skeletal model, denoted as $T=\{C_1',C_7',T_7',L_4',L_4'\}$. Landmarks are matched one-to-one: $s_i \leftrightarrow t_i$, where $i = 1,2,3,4,5$, $s_i\in S, t_i\in T$. The relationship between them is defined as in~(\ref{match_point}), where $l \textgreater 0$ is the scaling factor, $R\in SO(3)$ is the 3D rotation matrix. $f\in R^3$ is the 3D translation vector. $\varepsilon_i\in R^3$ denotes the unknown additive noise and assumes that the noise obeys a zero-mean isotropic Gaussian distribution with a standard deviation of $\sigma _i$.
\begin{equation}
    t_i=l\cdot R\cdot s_i+f+\varepsilon  _i,i=1,2,3,4,5
    \label{match_point}
\end{equation}

\par Under the condition of obeying the maximum likelihood estimation, the optimization solves for the scaling factor $l^*$, the rotation matrix $R^*$ and the translation vector $f^*$, as in~(\ref{optimization_solve}).
\begin{equation}
    l^{*}, R^{*}, f^{*}=\arg \min _{l>0, R \in \operatorname{SO}(3), t \in \mathbb{R}^{3}} \sum_{i=1}^{5} \frac{1}{\sigma_{i}^{2}}\left\|t_{i}-l R s_{i}-f\right\|_{2}^{2}
    \label{optimization_solve}
\end{equation}
\par The combination of spatial transformations $(l^*,R^*,t^*)$ obtained from the optimization solution is applied to the spine model reconstructed from the medical images taken before the operation to obtain the preliminary alignment.

\par Next, we further use the ARAP algorithm for exact alignment to translate and transform the pelvis and sacrum positions in the 3D model of the patient's spine to align them to the corresponding positions in the body surface skeletal model.
\par Specifically, we aim to translate the pelvis and sacrum in the 3D model of the patient's spine to the corresponding positions in the body surface skeletal model, while maintaining the rigidity of the other parts. To this end, an initial displacement is set to translate the pelvis and sacrum to the target position. Remember that the initial position of the pelvis and sacrum in the 3D model of the patient's spine is $p_i$ and the position in the target body surface skeletal model is $v_i$.
Next, for each vertex $v_i$, the set $N(i)$ of its neighboring vertices is found. For each vertex $v_j \in N(i)$, the local rigidity transformation in the ideal case is computed. Then, the energy function is constructed, as in~(\ref{energy_function}):
\begin{equation}
    E(v)=\sum_{i\in V }^{} \sum_{j\in N(i)}^{} ||(v_i-v_j)-R_i(p_i-p_j)||^2,
    \label{energy_function}
\end{equation}
where $R_i$ is the local rigid transformation matrix, approximating a rotation matrix for vertex i. To maintain rigidity throughout the deformation process, we solve the linear system as in~(\ref{linear_system}).
\begin{equation}
    v_i=\frac{1}{N(i)}\sum_{j\in N(i)}^{}(R_ip_j+(v_j-R_jp_j)) 
    \label{linear_system}
\end{equation}

\begin{figure}[htbp]
    \centering
        \subfigure[Fusion with standard bone]{  % 第一个子图，xx 为第一个字体的名称
        \includegraphics[width=0.245\textwidth]{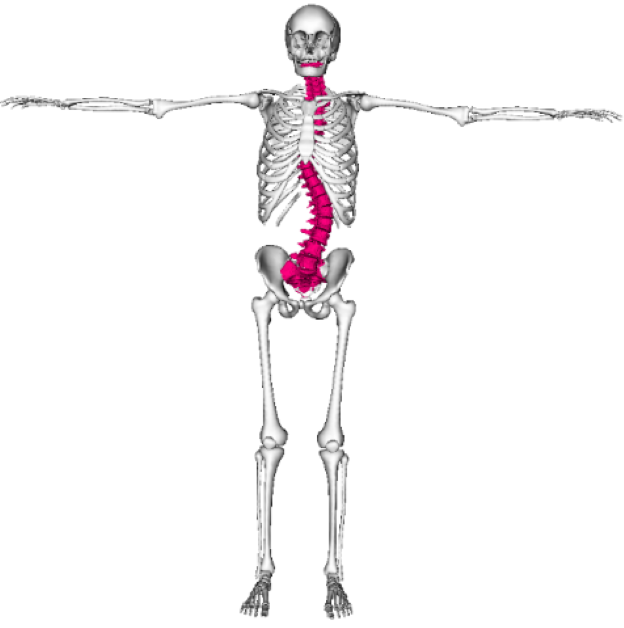}  % 图片路径和大小
}
        \subfigure[Fusion with SMPL model ]{  % 第二个子图
        \includegraphics[width=0.17\textwidth]{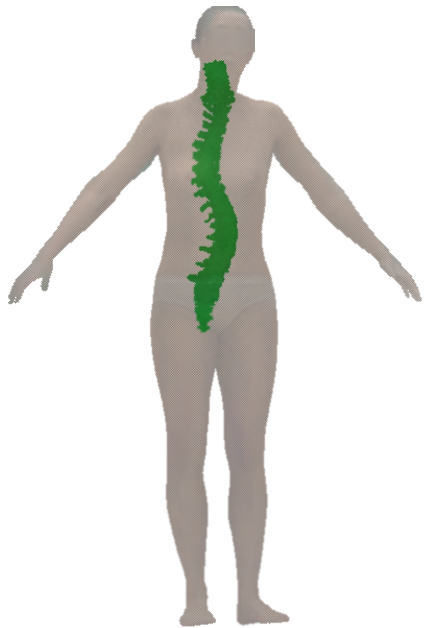}
}
    \caption{Depiction of the fused and aligned preoperative medical images. (a) The red spinal column, reconstructed from CT, represents the patient's true spinal anatomy, with the white skeletal framework being the standard bone model derived from the body surface, illustrating the registered outcome. (b) The human skin model superimposed on the result of (a).}    % 整个大图的标题
    \label{result}
\end{figure}

The method ensures an accurate match between the patient-specific spine model and the body skeleton model in terms of geometry and relative position, thus enabling individualized spine alignment and reconstruction. Eventually, an accurate spine model after fusion is obtained, as shown in Fig.~\ref{result}.

\section{Experiment} \label{Sec:experiment}
\subsection{Implement Detail}
\textbf{Data collection and pre-processing}:
In this study, we collected experimental data from six IS patients from the Department of Orthopedics, Peking Union Medical College Hospital, including:
 \par1) Multi-view recordings of T-positions: These images were utilized to derive a body surface model for each subject.
 \par 2) CT images of the lumbar spine in the supine position: Each patient underwent a CT scan capturing the anatomy of the lumbar spine.
 \par3) X-rays of the spine and corresponding image data: These were taken with the patient in an upright position.

\textbf{Experiment detail}:
In order to obtain the human body point cloud data, we firstly estimated the SMPL model on the recorded patient images. The SMPL model provided a standard 3D human body model for representing the posture and shape of the human body. Next, we fed the UV maps generated by the SMPL model (with a resolution of 128×128×3) into a pose encoder. This encoder utilized the standard U-Net architecture and consisted of five [Conv2d, BatchNorm, LeakyReLU] blocks, followed by five [ReLU, ConvTranspose2d, BatchNorm] blocks, with the BatchNorm omitted in the last block.
\par In training, an optimizable feature tensor with the same resolution as the output of the pose encoder (128 × 128 × 64) was trained using an automatic decoding method. The output of the pose encoder was then combined with the optimized feature tensor and fed into a Gaussian parameter decoder. To obtain finer details, the combined feature tensor was 4× upsampled to obtain a dimension of 512×512×64. The final output 3D Gaussian point cloud contained about 200,000 points.
\par In the process of predicting the standard human skeletal model, we used a pre-trained model trained on a DXA (dual-energy X-ray absorptiometry) image dataset of 1000 subjects. The model focused on the regression from body shape parameters to skeleton shape parameters, a step that was critical to accurately represent the shape of the human skeleton. By utilizing the pre-trained model, robust and reliable initial estimates were provided, which were essential for accurate bone fitting and further analysis.
\begin{figure}[htbp]
    \centering
        \includegraphics[width=\linewidth]{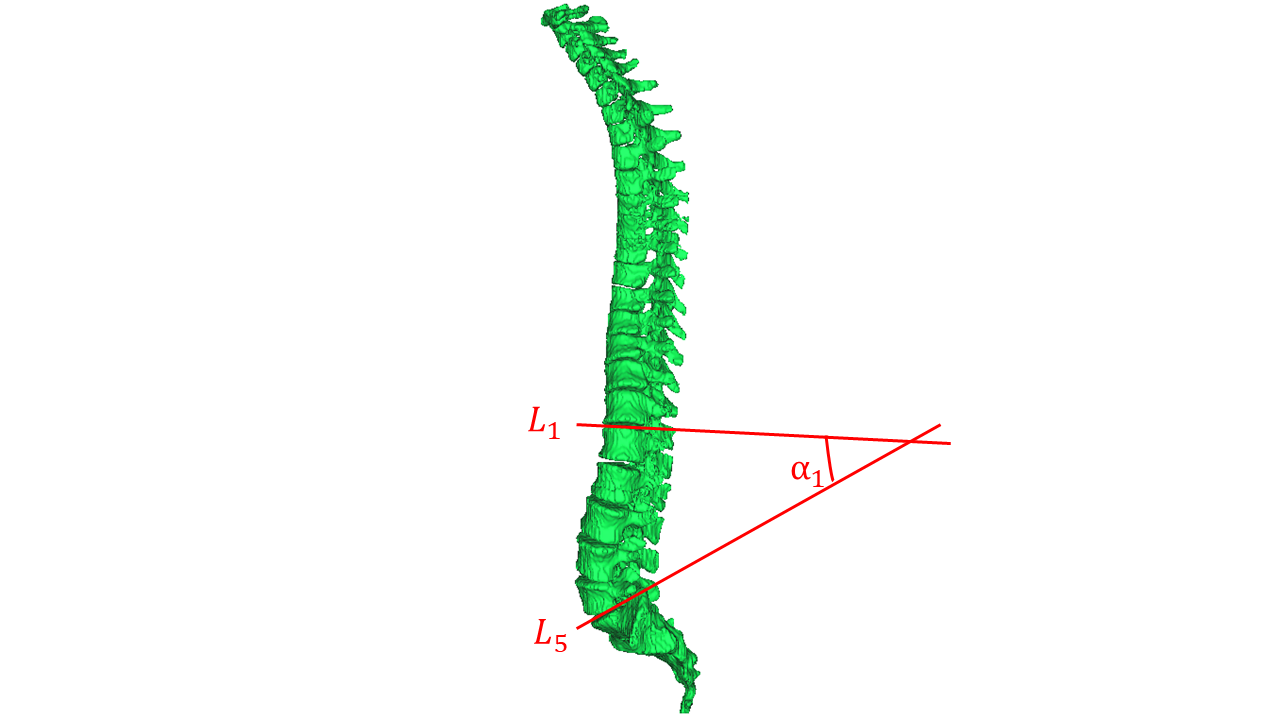}
    \caption{Definitions of Cobb angle}
    \label{RAsag}
\end{figure}

\subsection{Validation Metrics}
To validate the accuracy of the model in 3D space, we chose the the lumbar anterior convexity angle (measured by Cobb's angle method) as accuracy indicators to assess the error between the personalized skeletal model and the X-rays \cite{carlson2013comparison, di2013apical}.
The Cobb angle is one of the most frequently used methods in clinical practice to assess the severity of scoliosis. It helps clinicians determine the appropriate treatment strategy, such as whether bracing or surgery is required based on the measured angle. The lumbar anterior convexity angle, measured using the Cobb's angle method in the sagittal plane, is defined as the angle subtended by the line connecting the superior border of the L1 vertebrae to the inferior border of the L5 vertebrae, with posterior convexity defined as a positive value and anterior convexity as a negative value, denoted as $\alpha_1$, as shown in Fig.~\ref{RAsag}. By calculating the discrepancy between the X-ray measurements and the personalized skeletal models, we are able to assess the accuracy of the personalized skeletal models in reproducing the real anatomical structures observed in the X-ray images.
\begin{figure*}
	\centering
     \includegraphics[width=\textwidth]{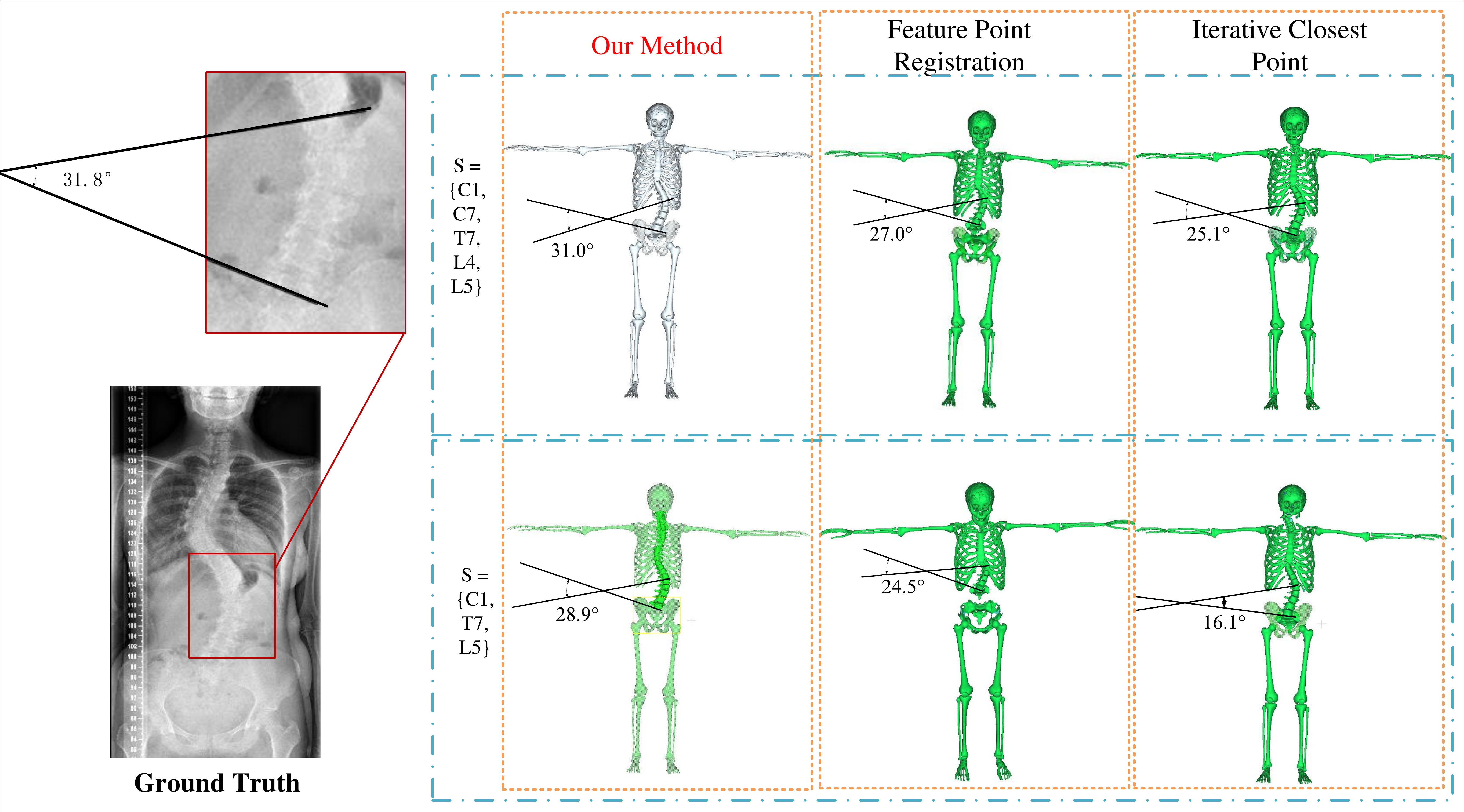}
	\caption{In this experiment, a comparative analysis was carried out on the Cobb angle values calculated on X-ray films and those obtained through different registration methods and feature point combinations, so as to intuitively present the differences among various methods in the construction of personalized bone models and the aspect of accuracy evaluation. }
    \label{Experiment}
\end{figure*}

\subsection{Results}
In the field of model reconstruction, CT images are usually acquired when patients are in the supine position. However, the personalized bone models generated are constructed based on the data of patients in the standard T-pose. Due to this difference in patient postures, special comparison strategies are urgently needed when evaluating the accuracy of the models. In view of this, when verifying the accuracy of personalized bone models, this study selects the patient X-ray images taken in the T-pose as the key reference basis. This method can effectively ensure the accuracy and consistency of the models in specific standard postures by conducting direct comparisons with X-ray data.
\par The main reason why the T-pose is selected as the standard pose is that it can construct a unified and reproducible reference framework for the model verification process, thus greatly eliminating the anatomical structure variations caused by different postures. In the T-pose, the anatomical distortion caused by changes in patient positions is minimized, which significantly improves the accuracy of the correspondence between X-ray images and the generated bone models in 3D space. In addition, X-ray imaging has the characteristic of high spatial resolution, which can accurately capture the structural changes of the spine under different loading and posture conditions. Based on these advantages, X-ray imaging has become an indispensable and important tool for evaluating the accuracy of bone models in the process of posture reconstruction.
\par This study has carried out a comparative experiment aiming to deeply explore the impact of different registration methods on model accuracy. Specifically, the Cobb angles are calculated on X-ray images and static digital human models constructed by different registration methods respectively. The registration methods adopted include the feature point registration combined with the ARAP method, the simple feature point registration method, and the Iterative Closest Point (ICP) registration combined with the ARAP method. The registration operations were performed under two different sets of feature point combinations {(C1, C7, T7, L4, L5), (C1, T7, L5)}. Through systematic comparison and analysis of the finally calculated Cobb angles, it is found that under the feature points (C1, C7, T7, L4, L5) selected in this study, the registration effect of the model constructed by the feature point matching combined with the ARAP method is the closest to the Cobb angle on the X-ray image. This result provides valuable reference for relevant research in the fields of model reconstruction and accuracy evaluation and helps to promote further technological development and optimization in these fields, as shown in Fig.~\ref{Experiment}, and Table \ref{tab:angle_comparison} presents the experimental comparison between six patients using different methods, demonstrating that our approach achieves the smallest error compared with ground truth, with a deviation less thanc.

\begin{table*}[htbp]
    \centering
    \caption{Comparison of Cobb angle measurements between methods and ground truth}
    \label{tab:angle_comparison}
    \resizebox{\textwidth}{!}{%
    \begin{tabular}{cccccccc}
    \toprule
     \multirow{2}{*}{Patient} & \multirow{2}{*}{Ground Truth } & \multicolumn{3}{c}{$S = ({C_1, C_7, T_7, L_4, L_5})$} & \multicolumn{3}{c}{ $S' = ({C_1, T_7,  L_5})$ } \\
    \cmidrule(lr){3-5} \cmidrule(lr){6-8}
     &  & Our Method & Feature Point Registration & Iterative Closest Point & Our Method & Feature Point Registration & Iterative Closest Point \\
    \midrule
    Patient1 & $31.8\degree$ & $31.0\degree$ & $27.0\degree$ & $25.1\degree$ & $28.9\degree$ & $24.5\degree$ & $16.1\degree$ \\
    Patient2 & $25.6\degree$ & $26.2\degree$ & $23.6\degree$ & $22.3\degree$ & $27.3\degree$ & $22.4\degree$ & $21.5\degree$ \\
    Patient3 & $22.5\degree$ & $22.0\degree$ & $21.2\degree$ & $25.2\degree$ & $22.2\degree$ & $21.5\degree$ & $24.1\degree$ \\
    Patient4 & $38.2\degree$ & $37.5\degree$ & $33.2\degree$ & $30.8\degree$ & $35.8\degree$ & $30.5\degree$ & $23.8\degree$ \\
    Patient5 & $33.5\degree$ & $32.5\degree$ & $29.1\degree$ & $26.7\degree$ & $31.0\degree$ & $29.5\degree$ & $20.6\degree$ \\
    Patient6 & $28.1\degree$ & $28.2\degree$ & $25.8\degree$ & $23.0\degree$ & $26.7\degree$ & $23.5\degree$ & $20.9\degree$ \\
    \bottomrule
    \end{tabular}%
    }
\end{table*}

%%%%%%%%%%%%%%%%%%%%%%%%%%%%%%%%%%%%%%%%%%%%%%%%%%%%%%%%%%%%%%%%%%%%%

\section{Conclusion and future works}
In this study, we establish an accurate static digital human model that uniquely integrates spinal characteristics specific to AIS patients. Unlike traditional imaging approaches, our method utilizes a novel combination of multi-view image data and CT-derived spine reconstructions. By leveraging advanced registration techniques, including feature-point alignment combined with ARAP, our approach achieves superior precision, with Cobb angle deviations consistently under $1\degree$. This framework provides a robust foundation for future dynamic digital human modeling and applications in clinical diagnosis, surgical planning, and rehabilitation strategies.
The innovations and comparative advantages of our framework include:
\begin{itemize}
    \item\textbf{Precise Spine-Body Alignment}: By integrating multi-modal data, our model achieves seamless alignment between the reconstructed spine and the standard skeletal model, which is critical for accurate representation of AIS-specific spinal deformities.
    \item\textbf{Error Reduction in Clinical Metrics}: Our approach ensures a higher fidelity representation of spinal curvature and rotation compared to existing static modeling techniques, demonstrating reductions in Cobb angle measurement errors by up to $5\degree$, when using ICP-based methods.
    \item\textbf{Clinical Utility and Scalability}: Beyond static modeling, this framework lays the groundwork for dynamic simulations and personalized interventions, providing clinicians with a more precise tool for surgical planning and rehabilitation strategies.
\end{itemize}
Future research will focus on the following directions:
\begin{itemize}
    \item \textbf{Dynamic Digital Human Development}: Expand upon the static model by integrating biomechanical principles and advanced motion simulation algorithms to simulate spine behavior during dynamic activities.
    \item \textbf{Broader Validation Across Demographics}: Test the method's applicability to diverse body types, age groups, and activity states, ensuring its reliability and generalizability.
    \item \textbf{Clinical Data Integration}: Incorporate longitudinal clinical data, including patient histories and outcomes, to enhance the model’s relevance and predictive capacity in personalized medicine.
\end{itemize}
Our study represents a significant step forward in the field of AIS modeling, addressing critical limitations in current methods and paving the way for dynamic digital human applications in clinical and research settings.

\section{Acknowledgment}
This work was supported in part by the National High Level Hospital Clinical Research Funding under Grant No. 2022-PUMCH-B-002, in part by the Beijing Nova Program under Grant No. 20230484488, and in part by the Beijing Natural Science Foundation under Grant No. L222012.
\textbf{Data availability} The datasets generated and/or analyzed during the current study are not publicly available due to patient privacy and confidentiality concerns
\section{Declarations} 
\textbf{Conflict of interest} The authors have no relevant financial interests to disclose.


\begin{refcontext}[sorting = none]
\printbibliography
\end{refcontext}

\end{document}